\documentstyle[rotate,graphicx]{mn}

\def\msun{{\rm M_{\odot}}}

\title[Ultraluminous X--ray Sources and Star Formation]
{Ultraluminous X--ray Sources and Star Formation}
\author[A.R.~King]{
A.R.~King\\
Theoretical Astrophysics Group, University of Leicester,
Leicester, LE1~7RH, UK}

\begin{document}

\maketitle

\begin{abstract}
{\it Chandra} observations of the Cartwheel galaxy reveal a population
of ultraluminous X--ray sources (ULXs) with lifetimes $\la 10^7$~yr
associated with a spreading wave of star formation which began some
$3\times 10^8$~yr ago. A population of high--mass X--ray binaries
provides a simple model: donor stars of initial masses $M_2 \ga
15\msun$ transfer mass on their thermal timescales to black holes of
masses $M_1 \ga 10\msun$.

For alternative explanations of the Cartwheel ULX population in terms
of intermediate--mass black holes (IMBH) accreting from massive stars,
the inferred production rate $\ga 10^{-6}$~yr$^{-1}$ implies at least
300 IMBHs, and more probably $3\times 10^4$, within the star--forming
ring. These estimates are increased by factors $\eta^{-1}$ if the
efficiency $\eta$ with which IMBHs find companions of $\ga 15\msun$
within $10^7$~yr is $<1$. Current models of IMBH production would
require a very large mass ($\ga 10^{10}\msun$) of stars to have
formed new clusters. Further, the accretion efficiency must be low
($\la 6\times 10^{-3}$) for IMBH binaries, suggesting super--Eddington
accretion, even though intermediate black hole masses are invoked with
the purpose of avoiding it.

These arguments suggest either that to make a ULX, an IMBH must
accrete from some as yet unknown non--stellar mass reservoir with very
specific properties, or that most if not all ULXs in star--forming
galaxies are high--mass X--ray binaries.

\end{abstract}
\begin{keywords}
accretion, accretion discs -- black hole physics --
X--rays: binaries -- stars: formation -- galaxies: starburst
\end{keywords}

\section{Introduction}

Recent observations of external galaxies have uncovered a population of
ultraluminous X--ray sources (ULXs: see Makishima et al., 2000 and
references therein) with apparent luminosities well above the
Eddington limit $L_{\rm Edd}$ for a stellar--mass black hole (or
neutron star). Several authors have suggested that ULXs reveal
accretion on to intermediate--mass black holes (IMBH), with masses
between stellar values and the $> 10^6\msun$ inferred for active
galactic nuclei (e.g. Colbert \& Mushotzky, 1999; Ebisuzaki et al.,
2001).

An alternative view (King et al., 2001) holds that ULXs are a bright,
shortlived, but common phase of stellar--mass X--ray binary
evolution. There is observational support for this, as Grimm, Gilfanov
\& Sunyaev (2002; see Table 1) show that apparently super--Eddington
episodes are not uncommon in stellar--mass binaries. For example the
high--mass system V4641 Sgr had a luminosity in the Chandra X--ray
band of $3.3\times 10^{39}$erg s$^{-1}$, despite a measured black hole
mass of $9.6\msun$. The high apparent luminosities of the ULXs may
either result from viewing a significantly anisotropic radiation
pattern at a favourable angle, or be genuinely super--Eddington (cf
Shaviv 1998, 2000; Begelman, 2002) or both. This form of `beaming'
need not involve relativistic effects, although Markoff et al., (2001)
and Koerding et al. (2002) have suggested that Doppler boosting in a
relativistic jet could explain the high luminosities of ULXs.

A major clue to the nature of ULXs comes from the discovery of 7 ULXs
in {\it Chandra} observations of the Antennae (Fabbiano et al.,
2001). These strongly indicate a connection with recent massive star
formation. IMBH models for ULXs (Miller \& Hamilton, 2002; G\"urkan,
Freitag \& Rasio, 2003) incorporate this by considering dense star
clusters. These provide promising sites both for IMBH formation, and
for capturing stellar companions to provide the accretion source. The
X--ray binary picture (King et al., 2001) has a natural connection
with massive star formation since it invokes a phase of high--mass
X--ray (HMXB) binary evolution. This is the thermal--timescale mass
transfer that must follow the familar wind--fed HMXB phase, as is
probably seen in SS433 (King, Taam \& Begelman, 2000). (Note that ULXs
are also seen in elliptical galaxies; in the X--ray binary picture
these are bright, long--lasting outbursts of soft X--ray transients
such as GRS~1915+105, cf King, 2002.)

The spectacular recent {\it Chandra} observation of the Cartwheel
galaxy (Gao et al., 2003) gives the most graphic illustration so far
of the connection between ULXs and star formation. The Cartwheel is
well known as a site of recent massive star formation. Most of this is
in a crisp ring expanding about the point where an intruder galaxy
plunged through the gas--rich disc of the galaxy about $3\times
10^8$~yr ago. The Chandra image reveals more than 20 ULXs (defined as
$L_{\rm 0.5-10keV} \ga 3\times 10^{39}$~erg~s$^{-1}$). Most of these
(about 80\% of the entire X--ray emission from the galaxy) are in the
dominant star--forming sites located precisely in the southern
quadrant of the ring. Each source is brighter ($L_{\rm 0.5-10keV} \ga
6\times 10^{39}$~erg~s$^{-1}$) than the most luminous ULXs seen in the
Antennae. The lack of radial spread in these source positions,
together with the known expansion velocity of the ring, show that
these ULXs must have ages $\la 10^7$~yr (Gao et al., 2003). 

These observations clearly place very tight constraints on models for
ULXs. I examine these below.

\section{ULXs and star formation}

King et al. (2001) examined the statistics of ULX formation,
independently of the adopted model. If $n$ is the number of
currently--observed ULXs in some region of a galaxy, they showed first
that the total number of potentially active ULXs in this region is
actually
\begin{equation}
N = {n\over bd}
\label{1}
\end{equation}
where $b \leq 1$ is the beaming (radiation anisotropy) factor, and $d
\leq 1$ the duty cycle. If for example $bd << 1$, most ULXs would
either be radiating in directions away from our line of sight, or
currently in low states. In the Cartwheel we can say more: the dearth
of observed ULXs inside the expanding star--forming ring means that
most have now `died', i.e. ceased accreting. The total number of dead
ULXs inside the ring is thus
\begin{equation}
N_{\rm tot} = N{t_*\over t_{\rm life}} = {n\over bd}{t_*\over t_{\rm life}}
\ga {300\over bd}
\label{2}
\end{equation}
where $t_* \simeq 3\times 10^8$~yr is the time since the wave of star
formation began to propagate outwards, $t_{\rm life} \la 10^7$~yr
is the ULX lifetime in the ring, and I have used eq. (\ref{1}) with $n
\sim 10$. This estimate could be still larger if as usual, the gas
surface density in the pre--intrusion galaxy increased exponentially
towards the dynamical centre.

Second, King et al. (2001) showed that the mass--transfer lifetime
of a ULX is
\begin{equation}
\tau = 10^6{m_2a\over bdL_{40}}~{\rm yr}.
\label{3}
\end{equation}
Here $m_2$ is the initial mass (in $\msun$) of the reservoir from
which the compact object accretes, $a \leq 1$ is the acceptance rate,
i.e. the fraction of transferred reservoir mass actually gained by the
accretor, and $L_{40}$ is the apparent (isotropic) bolometric
luminosity of the ULX in units of $10^{40}$~erg~s$^{-1}$. Dividing
(\ref{3}) into (\ref{1}) gives the important result that the required
birthrate of ULXs is independent of both anisotropy $b$ and duty cycle
$d$ (King et al., 2001). The Cartwheel observations now imply a
further pair of constraints. Clearly, one of these is that the
mass--transfer lifetime must be smaller than the inferred ULX
lifetime, i.e. $\tau < t_{\rm life}$. Using (\ref{3}) this gives
\begin{equation}
m_2 < 10 {bd\over a}L_{40}.
\label{4}
\end{equation}
If the mass reservoir for the ULX is a binary companion star, we must
also require that its main--sequence lifetime is less than $t_{\rm
life}$. Otherwise ULXs with initially wide separations would start
mass transfer only long after the wave of star formation had
passed. This would lead to a roughly uniform distribution of ULXs
inside the ring, quite unlike the sharp concentration at the ring edge
actually observed. This constraint translates directly into a limit on
the initial companion mass
\begin{equation}
m_2 \ga 15
\label{5}
\end{equation}
e.g. Iben (1967). With (\ref{4}) this gives
\begin{equation}
a < 0.6bdL_{40}
\label{6}
\end{equation}
This inequality must be satisfied by a comfortable margin to avoid the
requirement that all ULXs should form with a narrow range of companion
masses near $15\msun$. We can now apply the constraints (\ref{2}, \ref{5}
\ref{6}) to the two types of models for ULXs.

\subsection{Stellar--mass ULXs}

If the ULXs in the Cartwheel are stellar--mass binaries obeying the
Eddington limit we must have $b \la 0.1$. Constraint (\ref{5}) tells
us that these systems must be HMXBs, as expected. This probably means
that $d \sim 1$, so that (\ref{2}) implies 
\begin{equation}
N_{\rm tot} \ga 3000.
\label{7}
\end{equation} 
This is quite reasonable for a population of HMXBs. Finally
(\ref{6}) gives 
\begin{equation}
a \la 0.06, 
\label{8}
\end{equation}
i.e. the accretion process must be very
inefficient, with most of the mass lost by the companion failing to
accrete on to the compact object.  This requirement is easily
satisfied for thermal--timescale mass transfer: the mass--loss rate
from the companion is $-\dot M_2 \sim M_t/t_{\rm KH} \ga
10^{-5}\msun$~yr$^{-1}$, giving $a \la 0.01$ for a $10\msun$ black
hole accretor if the Eddington limit applies.

\subsection{Intermediate--mass black hole ULXs}

If the Cartwheel ULXs contain IMBHs accreting from massive stars we
get a different set of constraints. Even without any assumptions about
radiation anisotropy or duty cycle, (\ref{2}) requires at least
$N_{\rm tot} = 300$ inactive IMBH within the star--forming
ring. However Kalogera et al (2003; see also King et al., 2001) show
that all such IMBH binaries are likely to be transient, with accretion
discs subject to the standard thermal--viscous instability. This by
definition means that the duty cycle $d < 1$. Disc theory does not yet
provide quantitative estimates of $d$; however observed disc--unstable
systems have $d \la 10^{-2}$, and there is considerable observational
evidence (Ritter \& King, 2001) to suggest that long--period systems
with large discs, as in these ULXs binaries, have even smaller duty
cycles $d \la 10^{-2} - 10^{-3}$. We thus get from (\ref{2}) the
constraint
\begin{equation}
N_{\rm tot} \ga {3\times 10^4\over bd_{-2}}
\label{9}
\end{equation}
where $d_{-2}$ is $d$ in units of $10^{-2}$. From (\ref{6}) we get
\begin{equation}
a < 6\times 10^{-3}bd_{-2}.
\label{10}
\end{equation}

\section{Discussion}

I have considered explanations of the ULXs of the Cartwheel in terms
of stellar--mass and IMBH binaries. Subsection 2.1 suggests that a
population of high--mass X--ray binaries offers a reasonable picture.
These systems must accrete at super--Eddington rates, as expected in
the thermal--timescale mass transfer phase. These rates in turn
suggest possible lines of explanation for the high apparent
luminosities. The accretors in these HMXB binaries must be black
holes with typical masses $M_1 \ga 10\msun$, as Roche lobe overflow is
dynamically rather than thermally unstable for initial mass ratios
$M_1/M_2$ below some critical value (Webbink, 1977; Hjellming, 1989)
which is probably of order 0.5 for the massive donors ($M_2 \ga
15\msun$) in these HMXBs. (Dynamical--timescale mass transfer is ruled
out as it would extinguish the binaries as X--ray sources.) Black hole
accretors with $M_1 \ga 10\msun$ also satisfy the constraint that
their progenitors must have lifetimes $\la 10^7$~yr, whereas this is
probably not true of neutron stars for example.

Explanations of the Cartwheel ULXs invoking IMBHs accreting from
massive stars run into problems. The required production rate $\sim
10^{-6}$~yr$^{-1}$ of IMBH implies a minimum total $N_{\rm tot}$ of
$\ga 300$, or more probably $\ga 3\times 10^4$, within the
star--forming ring. Each of these IMBH must have found a stellar
partner of $\ga 15\msun$. If this process has efficiency $\eta < 1$
the above estimates of $N_{\rm tot}$ increase by factors $\eta^{-1}$,
i.e. to $N_{\rm tot} \ga 3\times 10^4{\eta}^{-1}$. Models of IMBH
formation in clusters (Miller \& Hamilton, 2002; G\"urkan, Freitag \&
Rasio, 2003) predict that only one IMBH is produced by a typical
cluster mass of $3\times 10^5\msun$. Hence if all the ULXs in the
Cartwheel are assumed to be IMBH this requires a mass $\ga
10^{10}{\eta}^{-1}\msun$ to have appeared in star clusters since the
intrusion event, which seems unlikely. Finally it appears that the
accretion efficiency $a$ must be low ($\la 6\times 10^{-3}$) for IMBH
binaries. Of course this is not implausible for transient outbursts in
which the accretion rate is very high; however it is perhaps
disappointing to find super--Eddington accretion rates reappearing in
a model specifically designed to exclude them.

Given these results, one should consider ways of rescuing the IMBH
idea. There appear to be three main possibilities.

(a) Conditions at the current position of the star--forming ring may
be highly unrepresentative of the region within it. This idea lacks
plausibility, and would be completely ruled out if similar results are
found in other galaxies.

(b) IMBH do not accrete from stars to make ULXs in starburst galaxies,
but from some other kind of mass reservoir. There is no obvious
candidate for this reservoir, and the constraints that accretion must
should occur at rates $\ga 10^{-6}\msun$~yr$^{-1}$ and shut off after
$\sim 10^7$~yr are severe.

(c) Most if not all of the ULXs found in regions of star formation are
indeed HMXBs. However there is currently no clinching argument against
a small minority containing IMBHs, as all the arguments presented here
refer to population rather than individual source properties.

This last idea appears the most plausible. It is supported by the work
of Grimm, Gilfanov \& Sunyaev (2003), who show that ULXs fit on to the
X--ray luminosity functions of nearby star--forming galaxies when
these are normalised by the star formation rate.

\section{Acknowledgments}

I thank Martin Beer, Hans Ritter, Dan Rolfe and Klaus Schenker for
helpful discussions. Research in theoretical astrophysics at Leicester
is supported by a PPARC rolling grant. I gratefully acknowledge a
Royal Society Wolfson Research Merit Award.

\end{document}